\documentclass[conference]{IEEEtran}
\IEEEoverridecommandlockouts
\usepackage{cite}
\usepackage{amsmath,amssymb,amsfonts}
\usepackage{algorithmic}
\usepackage{graphicx}
\usepackage{textcomp}
\usepackage{xcolor}

\def\BibTeX{{\rm B\kern-.05em{\sc i\kern-.025em b}\kern-.08em
    T\kern-.1667em\lower.7ex\hbox{E}\kern-.125emX}}

\usepackage{amsthm}
\usepackage{balance}
\usepackage{dblfloatfix}
\usepackage{hyperref}

\newtheorem{theorem}{Theorem}

\begin{document}

\title{Quantum Circuit Pruning: From NISQ Architectures to Fault-Tolerant Operations
\thanks{
This work was supported by the European Commission (QUADRATURE: 101099697, WINC: 101042080), the MICIU and European ERDF (MCIN/AEI/10.13039/501100011033, PID2021123627OB-C51, PID2024-158682OB-C31, QCOMM-CAT), Generalitat Valenciana (CIPROM/2022/66), NextGenerationEU (PRTR-C17.I1), and FPI-UPC grant (Banco Santander).
}
}

\author{
    \IEEEauthorblockN{
        Pau Escofet\IEEEauthorrefmark{1},\
        Carmen G. Almudéver\IEEEauthorrefmark{2},\
        Sergi Abadal\IEEEauthorrefmark{1},\ and
        Eduard Alarcón\IEEEauthorrefmark{1}
    }
    \IEEEauthorblockA{
        \textit{\IEEEauthorrefmark{1}Universitat Politècnica de Catalunya, Spain} \qquad\quad
        \textit{\IEEEauthorrefmark{2}Universitat Politècnica de València, Spain}\\
        pau.escofet@upc.edu
        \vspace{-15pt}
    }
}

\maketitle

\begin{abstract}
We present a routing-aware pruning strategy for quantum circuits that selectively removes parametric two-qubit gates whose computational contribution is outweighed by the fidelity cost of their execution, and assess it across both Noisy Intermediate-Scale Quantum (NISQ) and fault-tolerant quantum computing (FTQC) architectures. Each two-qubit gate of a circuit is evaluated by comparing an architecture-independent lower bound on the fidelity cost of omitting the gate against a hardware-specific model for the routing execution cost in each setting. We additionally provide an analytical characterization of the pruning boundary as a function of rotation angle, qubit distance, and hardware noise. Simulations on benchmark circuits demonstrate gate count reductions of up to 48.6\% and fidelity improvements of up to 47.7\% in the NISQ regime, while a feasibility analysis on FT hardware indicates that pruning is expected to remain relevant across a range of code distances and physical error rates. These results establish routing-aware pruning as a practical and unified compilation strategy across the quantum computing stack.
\end{abstract}

\begin{IEEEkeywords}
Quantum Circuit Mapping, NISQ, FTQC
\end{IEEEkeywords}

\section{Introduction}
Quantum computing holds the promise of solving problems that are intractable for classical computers, with potential breakthroughs in areas such as cryptography, optimization, chemistry, and machine learning \cite{shor1999polynomial}. While large-scale fault-tolerant quantum computers remain a long-term goal, significant progress is being made with noisy intermediate-scale devices, and the field is advancing towards error-corrected architectures \cite{eisert2025mind}. In both regimes, the reliability of computations is severely constrained by noise and the overhead introduced during circuit compilation. To maximize the utility of current and near-future hardware, a vast body of research has focused on improving circuit fidelity through noise mitigation, qubit mapping, and circuit optimization \cite{escofet2024route, siraichi2018qubit}.

A key source of compilation overhead in both Noisy Intermediate-Scale Quantum (NISQ) and Fault-Tolerant Quantum Computing (FTQC) architectures is the cost of executing two-qubit gates between non-adjacent qubits. On NISQ devices, this requires inserting extra operations that accumulate and propagate noise; on FT architectures, it usually requires establishing a merge-split path between logical qubits. In both cases, controlled gates with small rotation angles may offer less computational benefit than the routing overhead required for their execution, making them candidates for removal.

In this work, we introduce and extend a routing-aware pruning strategy that evaluates parametric two-qubit gates in terms of their expected impact on the quantum state relative to the routing cost required for their execution. Building on our prior work \cite{escofet2026quantum}, which established the NISQ case, we here extend the framework to fault-tolerant architectures based on lattice surgery, providing a unified pruning methodology applicable across hardware regimes. For the NISQ setting, we additionally provide an analytical characterization of the pruning boundary as a function of rotation angle, qubit distance, and hardware noise, offering principled insight into when and why pruning is beneficial. We assess the methodology on benchmark circuits compiled to grid-based NISQ architectures and on lattice surgery circuits under realistic noise models.

Results demonstrate that routing-aware pruning consistently reduces two-qubit gate counts (up to 48.6\%) and improves final state fidelity (up to 47.7\%) in the NISQ regime, with gains becoming more pronounced as circuit size and routing complexity increase. In the FT setting, the framework identifies meaningful pruning opportunities across several code distances and hardware noise regimes. These results highlight the potential of integrating routing-aware pruning into compilers targeting both near-term and future quantum hardware.

\section{Quantum Computing Architectures}

\subsection{Noisy Intermediate-Scale Quantum Processors}

Current NISQ processors suffer from imperfect gate operations and significant decoherence over time, degrading the quantum state fidelity and constraining computational capability \cite{preskill2018quantum, arute2019quantum}. One of the main additional constraints of current processors, particularly in superconducting platforms \cite{nakamura1999coherent}, is the limited connectivity between qubits. In this technology, physical qubits are only coupled to their immediate neighbors, rendering non-adjacent two-qubit gates infeasible \cite{arute2019quantum, google2025quantum}. The compiler, which adapts the circuit to a specific platform, maps the virtual qubits from the circuit to the physical qubits on the device, and \texttt{SWAP} gates are inserted to move qubits and bring them into neighboring positions whenever a two-qubit gate acts on distant qubits. We refer the reader to \cite{escofet2024route, siraichi2018qubit} for an in-depth discussion of quantum circuit mapping.

Every additional \texttt{SWAP} gate introduces further noise, and in deep circuits the fidelity loss from routing alone can dominate the total error budget. The impact is especially severe for circuits with many long-range interactions compiled onto sparsely connected devices. Several gate pruning strategies have been proposed to alleviate this, removing gates whose contribution to the quantum state is outweighed by the noise they introduce \cite{imamura2023offline, kulshrestha2024qadaprune}. However, existing approaches typically operate before routing and consider only the rotation angle of the gate, without accounting for the connectivity constraints of the target hardware. This gap motivates our routing-aware pruning strategy, which integrates pruning decisions directly into the compilation flow.

\subsection{Fault-Tolerant Quantum Computing Architectures}
Quantum error correction (QEC) offers a path beyond the limitations of NISQ hardware by encoding a single logical qubit across many physical qubits, distributing information such that local errors can be detected and corrected without disturbing the encoded state \cite{dennis2002topological}. Among the most promising QEC schemes is the surface code, which requires only nearest-neighbor interactions and achieves high fault-tolerance thresholds among other remarkable properties \cite{dennis2002topological, fowler2012surface}.

In surface code architectures (or other topological codes \cite{bombin2006topological}), logical qubits are patches of physical qubits on a planar layout, with logical Clifford gates implemented via lattice surgery (LS) \cite{horsman2012surface, litinski2019game}. LS performs two-qubit logical operations by temporarily merging adjacent logical patches along a shared boundary and then splitting them apart, as depicted in Figure \ref{fig:pruning_overview} c). We refer the reader to \cite{fowler2012surface, horsman2012surface, litinski2019game} for an in-depth explanation of the surface code and LS.

When the two logical patches involved in a gate are not adjacent, a path of ancilla patches must be established between them to mediate the merge-split sequence. The length of this path directly impacts the logical error rate (LER) of the operation (\textit{i.e.}, the failure probability of the gate), where longer paths expose the computation to more opportunities for error during the merge and split steps, making it a key factor in determining gate fidelity and a natural basis for pruning decisions analogous to those developed for the NISQ setting.

\section{Pruning Circuits: Improving Fidelity via Circuit Approximation}

A natural strategy for improving circuit fidelity on noisy hardware is to assess, for each gate in the quantum circuit, whether executing it actually contributes to the computation, or whether the noise it adds disrupts it. Most gates implement non-trivial transformations which are essential to the algorithm. But for parametric gates with small rotation angles, the answer is less obvious: the quantum state change introduced by a near-identity rotation may be smaller than the noise introduced by the hardware operations required to execute it. This is the core intuition behind circuit pruning.

A two-qubit parametric gate, which applies a $\theta$-rotation to the target qubit, is a natural pruning candidate when $\theta$ is small, since the gate then acts close to the identity and its omission leaves the quantum state largely undisturbed. To quantify the impact of each parametric gate, we first establish a lower bound on the fidelity cost of omitting such a gate \cite{escofet2026quantum}.
\begin{theorem}
Given a $\theta$-rotation to an arbitrary quantum state $|\psi\rangle$, the fidelity between the rotated and unrotated states is lower-bounded by:
\begin{equation}
     F(|\psi\rangle, R_{\hat{n}}(\theta)|\psi\rangle) \geq \cos^2 \left( \frac{\theta}{2} \right)
     \label{eq:theo_op}
\end{equation}
\label{theo:rotation}
\end{theorem}
The bound follows from expanding the rotation operator and considering the worst-case quantum state (\textit{i.e.}, the one most sensitive to the rotation axis). As expected, for small $\theta$ the bound approaches unity, confirming that near-identity rotations are suitable candidates for pruning, while for $\theta = \pi$ the bound reaches zero, corresponding to an orthogonal rotation that should never be pruned (see \cite{escofet2026quantum} for the full proof).

\begin{figure*}
    \centering
    \includegraphics[width=0.925\linewidth]{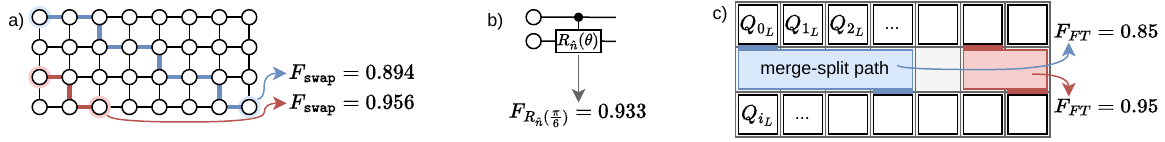}
    \caption{Routing-aware pruning decision for a parametric controlled rotation gate \textbf{(b)} in both the NISQ and FT settings. \textbf{(a)} NISQ case: the compiler evaluates the physical distance between the two qubits. The blue path is too long and the gate is pruned; the red path is short enough and the gate is retained. \textbf{(c)} FT case: logical qubits occupy the top and bottom rows, while the middle row contains ancilla patches used to mediate the merge-split path. The blue path is too long and the gate is pruned; the red path is short enough and the gate is retained. In both cases, the pruning threshold is set by the same bound $\cos^2 \left( \theta/2 \right)$.}
    \label{fig:pruning_overview}
\end{figure*}

This lower bound captures the worst-case fidelity cost of omitting the gate, depending only on $\theta$, making it architecture-independent. The pruning logic in both the NISQ and FTQC settings rests on comparing this omission cost against the execution cost imposed by the hardware, which we model separately for each regime. Figure~\ref{fig:pruning_overview} illustrates the pruning decision for both architectures: given a parametric gate (Figure~\ref{fig:pruning_overview} b), the compiler evaluates the routing path between the two qubits or logical patches involved (Figures~\ref{fig:pruning_overview} a and~\ref{fig:pruning_overview} c), and prunes the gate when the path is long enough so that 
its impact on fidelity is higher than omitting the gate.

\subsection{Pruning in NISQ Routing}
On NISQ hardware, executing a two-qubit gate between non-adjacent qubits requires a \texttt{SWAP} chain to bring them together. A \texttt{SWAP} gate decomposes into three CNOT gates, each introducing depolarizing noise $p_2$. The fidelity of executing a gate between qubits $q_i$ and $q_j$ separated by distance $d_{q_i, q_j}$ is:
\begin{equation}
    F_{\texttt{swap}} := \left( (1-p_2)^{3 \cdot d_\texttt{swap}} + \frac{1 - (1-p_2)^{3 \cdot d_\texttt{swap}}}{4} \right)^2
    \label{eq:f_swap}
\end{equation}
following the depolarizing noise fidelity model of \cite{escofet2025accurate}, where $d_\texttt{swap} = \left\lceil \frac{d_{q_i, q_j}}{2} \right\rceil$ is the number of \texttt{SWAP}s that each qubit must undergo to meet at the middle position.

Every time a parametric two-qubit gate is encountered during compilation, $F_{\text{swap}}$ is computed from the current physical positions of the two qubits, and compared against the pruning bound $\cos^2(\theta/2)$ from Theorem~\ref{theo:rotation}. If $F_{\text{swap}} < \cos^2(\theta/2)$, the routing overhead outweighs the computational contribution of the gate, and the gate is discarded. Crucially, this comparison is made \emph{during} routing rather than before it, so the decision reflects the actual qubit positions at compilation time rather than an abstract estimate of connectivity cost.

\begin{figure}
    \centering
    \includegraphics[width=1\linewidth]{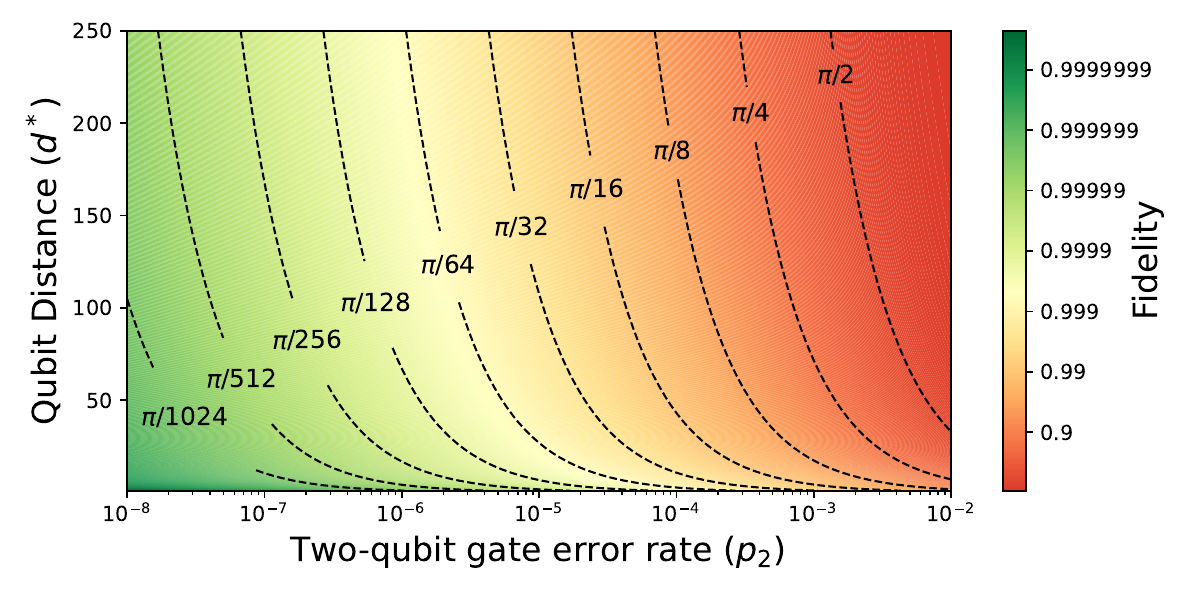}
    \caption{Analytical characterisation of the pruning boundary in the NISQ setting. The color map shows $F_{\text{swap}}$ as a function of the two-qubit gate error rate $p_2$ and the qubit routing distance. Dashed lines indicate the critical distance $d^*$ for parametric gates with angles from $\pi/2$ to $\pi/1024$. Gates whose interacting qubits are separated by a distance above the corresponding dashed line will be pruned, while those below will be maintained.}
    \label{fig:pruning_boundaries}
\end{figure}

To characterize the interplay between noise, rotation angle, and routing distance, Figure~\ref{fig:pruning_boundaries} shows, from a direct analytical evaluation of Eq. (\ref{eq:f_swap}) against Theorem \ref{theo:rotation}, the critical distance $d^*$ beyond which a gate is pruned, as a function of $p_2$, and for $\theta \in [\pi/1024, \pi/2]$. As expected, $d^*$ grows with $\theta$ and decreases with $p_2$: larger rotation angles tolerate longer, noisier paths, while lower-noise hardware allows longer paths before execution cost outweighs the gate's contribution. At current NISQ noise levels ($p_2 \sim 10^{-3}$), even significant angles like $\theta = \pi/8$ are pruned at $d^* \approx 20$, showing how aggressively routing overhead penalizes execution today. As $p_2$ improves toward $\sim10^{-6}$, only near-identity rotations ($\theta \lesssim \pi/256$) remain pruning candidates at distances $\leq 250$, suggesting the method grows more selective as noise decreases.

Qubit routing is NP-complete \cite{siraichi2018qubit}, so heuristic algorithms are inherently suboptimal, inserting more SWAPs than $d_{q_i,q_j}$. In \cite{escofet2024route}, Route-Forcing was shown to insert $\sim1.2–1.3\times$ more SWAPs than alternative approaches. Based on this, we approximate the routing overhead as $1.25 \cdot d_{q_i,q_j}$ whenever mapping a circuit with Route-Forcing, following \cite{escofet2026quantum}.

\subsection{Pruning in FTQC Routing}

\begin{figure*}
    \centering
    \includegraphics[width=1\linewidth]{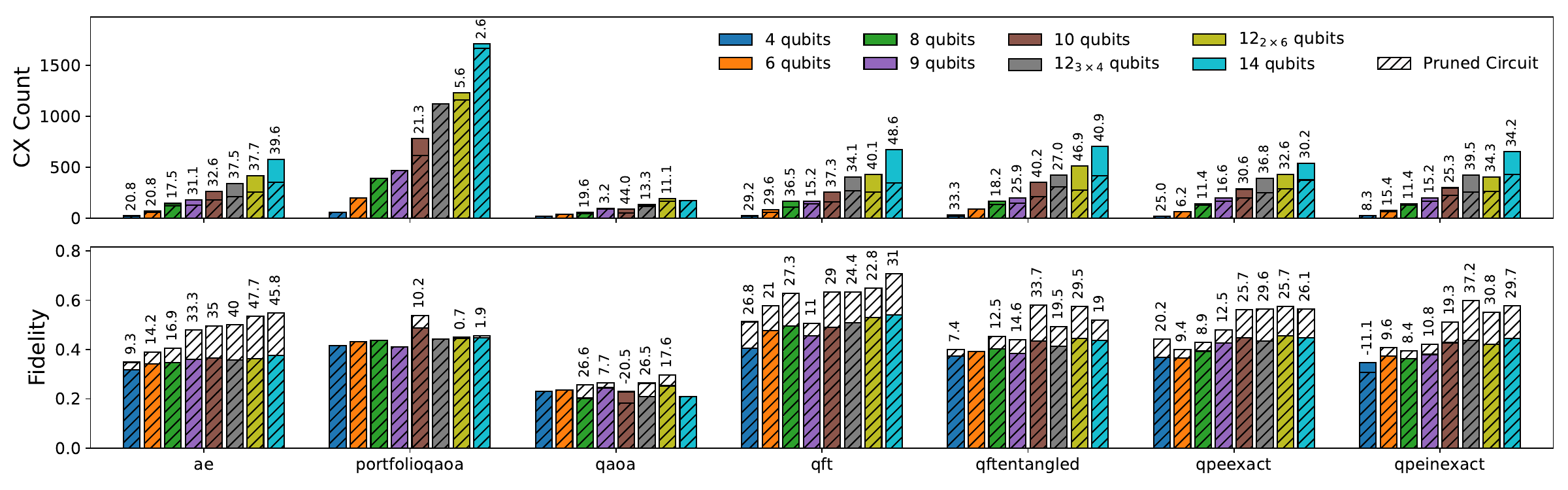}
    \caption{Two-qubit gate count \textbf{(top)} and circuit fidelity \textbf{(bottom)} for the selected circuits containing between 4 and 14 qubits, when compiled with (hatched bars) and without (solid bars) the proposed pruning strategy. The percentage improvement in fidelity achieved by pruning is indicated above each bar.}
    \label{fig:nisq_results}
\end{figure*}

\begin{figure*}
    \centering
    \includegraphics[width=1\linewidth]{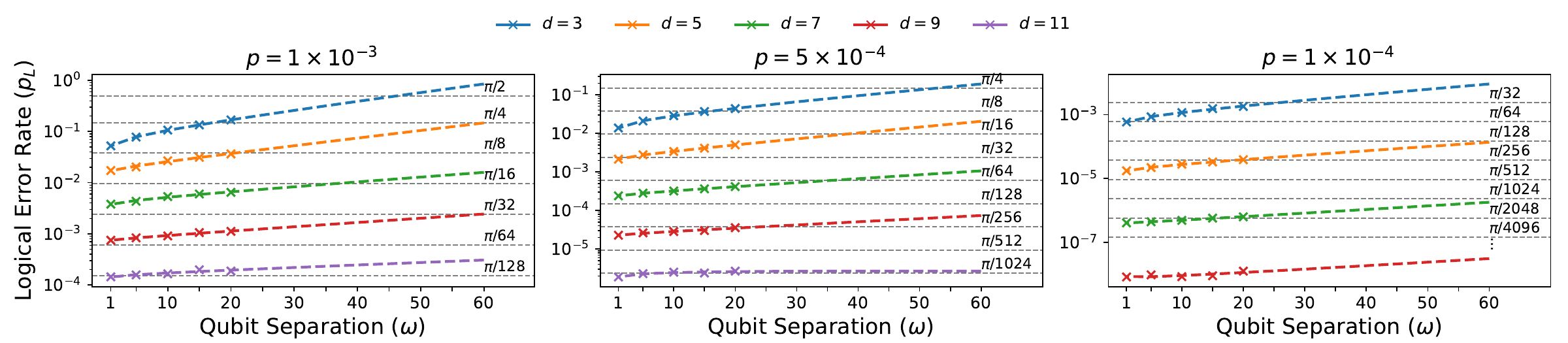}
    \caption{LER of two logical CNOT gates implemented via lattice surgery as a function of patch separation, for three depolarizing noise regimes and surface code distances $d_{\text{code}} \in \{3, 5, 7, 9, 11\}$. LER grows with patch separation as longer merge-split paths accumulate more errors. Horizontal dashed lines indicate the logical infidelity threshold $1 - \cos^2(\theta/2)$ for parametric gates with rotation angles $\theta$.}
    \label{fig:ftqc_results}
\end{figure*}

The same logic is used in the FT setting, but the cost model differs. Rather than depolarizing noise from \texttt{SWAP} insertions, execution overhead is determined by the LER of the merge-split path connecting the two logical patches.

The LER grows with the number of ancilla patches traversed, as each additional step introduces further exposure to errors during the merge-split operations \cite{litinski2019game}. We define a fidelity proxy for LS execution, $F_{\text{FT}} = 1-\text{LER}$, which decreases as the distance between logical patches increases.

The pruning criterion then mirrors the NISQ case: given a parametric controlled rotation of angle $\theta$, if $F_{\text{FT}}$ falls below $\cos^2(\theta/2)$, the routing overhead outweighs the computational contribution and the gate is pruned. Crucially, the lower bound on the omission fidelity established in Theorem~\ref{theo:rotation} is architecture-independent, making it applicable to both regimes without modification, with the only change between both regimes being the routing cost model.

It is worth noting that the methodology described here is not specific to any particular code. The same methodology applies to any QEC code and any approach for implementing logical two-qubit gates, including transversal gates. The only element that changes across settings is the model used to assess the fidelity cost of executing the gate ($F_{\text{FT}}$), which must be benchmarked for each specific code and gate implementation. In the following sections, we instantiate the framework for the concrete case of surface code with lattice surgery gate execution, and characterize the resulting pruning boundary through stabilizer circuit simulation \cite{gidney2021stim}.

\section{Results}

\subsection{Routing in NISQ}

We evaluate the proposed pruning methodology on a suite of benchmark circuits obtained from MQT Bench \cite{quetschlich2023mqt}, ranging from 4 to 14 qubits, compiled to grid topologies under a realistic noise model using Qiskit \cite{javadi2024quantum}. Fidelities are obtained via exact density-matrix simulation. Circuits are routed using the Route-Forcing algorithm \cite{escofet2024route} and decomposed into IBM's native gate set $\{\texttt{CX}, \texttt{ID}, \texttt{RZ}, \texttt{SX}, \texttt{X}\}$ after pruning.

For each circuit, three states are computed: the \emph{Ideal State}, obtained from a noiseless simulation; the \emph{Noisy State}, obtained by simulating the compiled circuit under the noise model and used as baseline; and the \emph{Pruned State}, generated with the same compilation and noise simulation procedure but with the proposed pruning strategy applied to parametric gates. Fidelity is assessed between the Ideal State and both the Noisy and Pruned States, an improvement in the latter fidelity will confirm that pruning alleviates compilation overhead.

The noise model, as described in \cite{escofet2026quantum}, assigns to each two-qubit gate a depolarizing noise parameter depending on the number of qubits and gates of the circuit, and thermal relaxation parameters $T_1$ and $T_2$ proportional to circuit duration. These values are chosen to produce a representative fidelity for the Noisy State rather than to reproduce a specific device, allowing the pruning improvement to be assessed cleanly against a controlled baseline.

Figure~\ref{fig:nisq_results} shows the number of $\texttt{CX}$ gates and circuit fidelity for all benchmarks and system sizes, when compiled with and without the proposed pruning strategy. For most circuits, pruning reduces the two-qubit gate count by up to 48.6\% while improving final state fidelity by up to 47.7\%, validating the core hypothesis that routing-aware pruning can improve fidelity and mitigate compilation overhead.

The benefit is particularly consistent for the Amplitude Estimation and QFT algorithms, where pruning improves fidelity across all circuit sizes. For circuits such as QAOA, improvements are less consistent across system sizes, with overly aggressive pruning at 10 qubits leading to a net fidelity degradation due to the fixed $1.25\times$ SWAP-overhead approximation underestimating routing efficiency for QAOA's dense structure, which allows the mapping algorithm to batch SWAP moves more effectively than assumed.

This methodology has been compared against offline pruning in \cite{escofet2026quantum}, where it matches the fidelity of the best offline approximation degree for the QFT circuit. Crucially, identifying this best degree requires exhaustively testing all approximation degrees beforehand, which is computationally infeasible for large-scale systems.

\subsection{Lattice Surgery Routing}

To evaluate the pruning framework in the FT setting, we benchmark the LER of LS operations as a function of patch separation using Stim \cite{gidney2021stim} and PyMatching \cite{higgott2022pymatching}, with up to $10^9$ shots per point. LS circuits are generated with TQEC \cite{suau2026tqec}.

Specifically, we benchmark the LER of two logical CNOT gates as a proxy for the merge-split path cost of a parametric two-qubit gate, since parametric two-qubit gates are not natively supported and must be decomposed into two CNOTs and single-qubit rotations \cite{kim2018efficient}. Beyond the cost of the two CNOTs required for a parametric gate, we omit the associated single-qubit rotations, since their execution does not depend on the distance between qubits. However, unlike in the NISQ case, these single-qubit rotations must be carried out via T-state injection following distillation. This process is currently the largest bottleneck in time and resources for FT estimates, and reducing the number of T-states required would substantially improve the resource cost of FT algorithms.

We apply a depolarizing noise model and explore three physical error rate regimes: $p \in \{10^{-3}, 5\times10^{-4}, 10^{-4}\}$, spanning the range from near-term fault-tolerant devices to more optimistic future hardware. For each noise value, we simulate surface codes with code distances $d_{\text{code}} \in \{3, 5, 7, 9, 11\}$ and vary the separation between logical patches ($\omega$).

It is worth noting that the two CNOTs constitute the sources of error in the decomposition that directly depend on the distance between qubits, since their LER  depends on how long is the merge-split path, while single-qubit rotations are applied locally and do not incur routing overhead. The reported two-CNOT LER therefore represents a lower bound on the true execution cost of the parametric gate.

Figure~\ref{fig:ftqc_results} reports the LER of two logical CNOTs as a function of patch separation for each noise regime and code distance. As expected, the LER grows with separation, since longer merge-split paths expose the computation to more error locations during the LS sequence. The horizontal dashed lines indicate the logical infidelity threshold $1 - \cos^2(\theta/2)$ for parametric rotations with angles $\theta \in \{\pi/4, \pi/8, \ldots, \pi/4096\}$: whenever the LER curve crosses above a dashed line, the gate with such angle $\theta$ would be pruned under our framework, since executing it introduces more logical error than omitting it.

At $p = 10^{-3}$, even moderate code distances like $d_{\text{code}} = 7$ exceed the pruning threshold for small-angle rotations at separations beyond a few tens of patches, indicating that pruning is relevant even for near-term FT devices. As $p$ decreases to $10^{-4}$, only very small rotation angles enter the pruning regime when increasing code distance, consistent with the expectation that better hardware narrows, but does not eliminate, the window where pruning is beneficial.

This analysis indicates where pruning is expected to be beneficial under realistic error regimes. An end-to-end application to full FT circuits remains to be conducted, as the required FT simulation tools are not currently available, and will likely have strong implications tied to magic state distillation.

\section{Conclusions}
We have introduced a routing-aware pruning methodology that selectively removes parametric two-qubit gates whose expected computational contribution is outweighed by the fidelity cost of their execution, applicable to both NISQ and fault-tolerant quantum computing architectures \cite{eisert2025mind}.

The framework rests on a single architecture-independent bound on the fidelity cost of omitting a parametric gate, and compares it against an execution cost model that is adapted to each hardware regime: noise accumulated through \texttt{SWAP} insertions in the NISQ case, and logical error rate along merge-split paths in the lattice surgery case.

For NISQ devices, simulations on benchmark circuits confirm that the method reduces two-qubit gate counts by up to 48.6\% and improves final state fidelity by up to 47.7\%, with improvements scaling with circuit size as routing overhead becomes the dominant source of error. An analytical characterization of the pruning boundary further reveals how the critical routing distance $d^*$ depends jointly on the rotation angle $\theta$ and the depolarizing noise parameter $p_2$, providing compiler designers with insights for anticipating pruning behavior across hardware platforms without requiring simulation.

For FT architectures, a feasibility analysis using Stim and TQEC indicates that pruning remains relevant even at moderate code distances and physical error rates, with the pruning window narrowing but persisting as hardware improves.

Taken together, these results establish routing-aware pruning as a practical and scalable compilation strategy across the quantum computing stack. The framework adapts naturally to the circuit structure and hardware connectivity of each architecture, delivering fidelity improvements without requiring prior tuning of approximation parameters.

\newpage
\bibliographystyle{IEEEtran}
\bibliography{IEEEabrv}

@article{shor1999polynomial,
  title={Polynomial-time algorithms for prime factorization and discrete logarithms on a quantum computer},
  author={Shor, Peter W},
  journal={SIAM review},
  volume={41},
  number={2},
  pages={303--332},
  year={1999},
  publisher={SIAM}
}

@article{preskill2018quantum,
  title={Quantum computing in the NISQ era and beyond},
  author={Preskill, John},
  journal={Quantum},
  volume={2},
  pages={79},
  year={2018},
  publisher={Verein zur F{\"o}rderung des Open Access Publizierens in den Quantenwissenschaften}
}

@article{arute2019quantum,
  title={Quantum supremacy using a programmable superconducting processor},
  author={Arute, Frank and Arya, Kunal and Babbush, Ryan and Bacon, Dave and Bardin, Joseph C and Barends, Rami and Biswas, Rupak and Boixo, Sergio and Brandao, Fernando GSL and Buell, David A and others},
  journal={nature},
  volume={574},
  number={7779},
  pages={505--510},
  year={2019},
  publisher={Nature Publishing Group UK London}
}

@article{google2025quantum,
  title={Quantum error correction below the surface code threshold},
  journal={Nature},
  volume={638},
  number={8052},
  pages={920--926},
  year={2025},
  publisher={Nature Publishing Group UK London}
}

@article{nakamura1999coherent,
  title={Coherent control of macroscopic quantum states in a single-Cooper-pair box},
  author={Nakamura, Yasunobu and Pashkin, Yu A and Tsai, JS},
  journal={nature},
  volume={398},
  number={6730},
  pages={786--788},
  year={1999},
  publisher={Nature Publishing Group UK London}
}

@inproceedings{escofet2024route,
  title={Route-forcing: Scalable quantum circuit mapping for scalable quantum computing architectures},
  author={Escofet, Pau and Gonzalvo, Alejandro and Alarc{\'o}n, Eduard and Almud{\'e}ver, Carmen G and Abadal, Sergi},
  booktitle={2024 IEEE QCE},
  volume={1},
  pages={909--920},
  year={2024},
  organization={IEEE}
}

@inproceedings{imamura2023offline,
  title={Offline quantum circuit pruning for quantum chemical calculations},
  author={Imamura, Satoshi and Kasagi, Akihiko and Yoshida, Eiji},
  booktitle={2023 IEEE QCE},
  volume={1},
  pages={349--355},
  year={2023},
  organization={IEEE}
}

@inproceedings{kulshrestha2024qadaprune,
  title={Qadaprune: Adaptive parameter pruning for training variational quantum circuits},
  author={Kulshrestha, Ankit and Liu, Xiaoyuan and Ushijima-Mwesigwa, Hayato and Bach, Bao and Safro, Ilya},
  booktitle={2024 IEEE QCE},
  volume={2},
  pages={120--125},
  year={2024},
  organization={IEEE}
}

@article{horsman2012surface,
  title={Surface code quantum computing by lattice surgery},
  author={Horsman, Dominic and Fowler, Austin G and Devitt, Simon and Meter, Rodney Van},
  journal={New Journal of Physics},
  volume={14},
  number={12},
  pages={123011},
  year={2012},
  publisher={IOP Publishing}
}

@article{litinski2019game,
  title={A game of surface codes: Large-scale quantum computing with lattice surgery},
  author={Litinski, Daniel},
  journal={Quantum},
  volume={3},
  pages={128},
  year={2019},
  publisher={Verein zur F{\"o}rderung des Open Access Publizierens in den Quantenwissenschaften}
}

@article{fowler2012surface,
  title={Surface codes: Towards practical large-scale quantum computation},
  author={Fowler, Austin G and Mariantoni, Matteo and Martinis, John M and Cleland, Andrew N},
  journal={Physical Review A—Atomic, Molecular, and Optical Physics},
  volume={86},
  number={3},
  pages={032324},
  year={2012},
  publisher={APS}
}

@inproceedings{escofet2026quantum,
  title={Quantum Circuit Pruning: Improving Fidelity via Compilation-Aware Circuit Approximation},
  author={Escofet, Pau and Rodrigo, Santiago and Sarkar, Rohit Sarma and Almud{\'e}ver, Carmen G and Alarc{\'o}n, Eduard and Abadal, Sergi},
  booktitle={2026 IEEE International Symposium on Circuits and Systems (ISCAS)},
  pages={1521--1525},
  year={2026},
  organization={IEEE}
}

@inproceedings{siraichi2018qubit,
  title={Qubit allocation},
  author={Siraichi, Marcos Yukio and Santos, Vin{\'\i}cius Fernandes dos and Collange, Caroline and Pereira, Fernando Magno Quint{\~a}o},
  booktitle={Proceedings of the 2018 international symposium on code generation and optimization},
  pages={113--125},
  year={2018}
}

@article{escofet2025accurate,
  title={An accurate and efficient analytic model of fidelity under depolarizing noise oriented to large scale quantum system design},
  author={Escofet, Pau and Rodrigo, Santiago and Garcia-S{\'a}ez, Artur and Alarc{\'o}n, Eduard and Abadal, Sergi and Almud{\'e}ver, Carmen G},
  journal={Quantum Science and Technology},
  volume={10},
  number={3},
  pages={035061},
  year={2025},
  publisher={IOP Publishing}
}

@article{javadi2024quantum,
  title={Quantum computing with Qiskit},
  author={Javadi-Abhari, Ali and Treinish, Matthew and Krsulich, Kevin and Wood, Christopher J and Lishman, Jake and Gacon, Julien and Martiel, Simon and Nation, Paul D and Bishop, Lev S and Cross, Andrew W and others},
  journal={arXiv preprint arXiv:2405.08810},
  year={2024}
}

@article{quetschlich2023mqt,
  title={MQT Bench: Benchmarking software and design automation tools for quantum computing},
  author={Quetschlich, Nils and Burgholzer, Lukas and Wille, Robert},
  journal={Quantum},
  volume={7},
  pages={1062},
  year={2023},
  publisher={Verein zur F{\"o}rderung des Open Access Publizierens in den Quantenwissenschaften}
}

@article{gidney2021stim,
  title={Stim: a fast stabilizer circuit simulator},
  author={Gidney, Craig},
  journal={Quantum},
  volume={5},
  pages={497},
  year={2021},
  publisher={Verein zur F{\"o}rderung des Open Access Publizierens in den Quantenwissenschaften}
}

@article{suau2026tqec,
  title={tqec: A Python package for topological quantum error correction},
  author={Suau, Adrien and Zhang, Yiming and Thakre, Purva and Zhao, Yilun and Dubey, Kabir and Bolanos, Jose A and Schelpe, Arabella and Hao, Tianyi and Seitz, Philip and Guerreschi, Gian Giacomo and others},
  journal={Journal of Open Source Software},
  volume={11},
  number={120},
  pages={9142},
  year={2026}
}

@article{higgott2022pymatching,
  title={Pymatching: A python package for decoding quantum codes with minimum-weight perfect matching},
  author={Higgott, Oscar},
  journal={ACM Transactions on Quantum Computing},
  volume={3},
  number={3},
  pages={1--16},
  year={2022},
  publisher={ACM New York, NY}
}

@article{bombin2006topological,
  title={Topological quantum distillation},
  author={Bombin, Hector and Martin-Delgado, Miguel Angel},
  journal={Physical review letters},
  volume={97},
  number={18},
  pages={180501},
  year={2006},
  publisher={APS}
}

@article{dennis2002topological,
  title={Topological quantum memory},
  author={Dennis, Eric and Kitaev, Alexei and Landahl, Andrew and Preskill, John},
  journal={Journal of Mathematical Physics},
  volume={43},
  number={9},
  pages={4452--4505},
  year={2002},
  publisher={American Institute of Physics}
}

@article{kim2018efficient,
  title={Efficient decomposition methods for controlled-R n using a single ancillary qubit},
  author={Kim, Taewan and Choi, Byung-Soo},
  journal={Scientific reports},
  volume={8},
  number={1},
  pages={5445},
  year={2018},
  publisher={Nature Publishing Group UK London}
}

@article{eisert2025mind,
  title={Mind the gaps: The fraught road to quantum advantage},
  author={Eisert, Jens and Preskill, John},
  journal={arXiv preprint arXiv:2510.19928},
  year={2025}
}

\end{document}